\documentclass{egpubl}
\usepackage{pg2026}

\SpecialIssuePaper         

\CGFStandardLicense

\usepackage[T1]{fontenc}
\usepackage{dfadobe}  
\usepackage{booktabs}
\usepackage{cite}  
\usepackage{amsmath,amssymb,amsfonts}
\BibtexOrBiblatex
\electronicVersion
\PrintedOrElectronic
\ifpdf \usepackage[pdftex]{graphicx} \pdfcompresslevel=9
\else \usepackage[dvips]{graphicx} \fi

\usepackage{egweblnk} 

\title[Telligram]%
      {Telligram: Text-Driven Calligram Generation via \\ Diffusion-Guided Skeleton Optimization}

\author[T. Shi \& P. Xu]{
\parbox{\textwidth}{\centering 
    Tianci Shi\orcid{0009-0009-3688-8471} and
    Pengfei Xu\thanks{Corresponding author: Pengfei Xu (xupengfei.cg@gmail.com)}\orcid{0000-0003-4770-4374}
        }
        \\
{\parbox{\textwidth}{\centering CSSE, Shenzhen University, China}
}
}

\begin{document}
\maketitle

\begin{abstract}
Compact calligram generation aims to form a semantic shape while keeping letters recognizable. Most existing methods are shape-conditioned and mainly solve downstream letter layout inside a given contour. We study text-only calligram generation without an input contour. This setting is difficult because semantic shape formation and letter readability strongly interfere with each other when optimized in a single stage. Pushing the word toward a clear figure can easily damage glyph structure, while preserving readable letters can weaken the target shape. To address this difficulty, we present Telligram, a training-free, low-tuning, two-stage framework composed of Semantic Occupancy Prior Formation and Readability-Constrained Glyph Realization. The first stage uses Variational Score Distillation (VSD) with structured skeleton optimization and hierarchical gradient projection to produce a semantic occupancy prior. The second stage converts this occupancy prior into per-letter regions and reconstructs readable glyph layouts through lightweight geometric processing. The framework generates coherent and creative word-level semantic calligrams directly from text prompts.

\end{abstract}

\section{Introduction}

Calligrams are a form of visual art in which letters or words are arranged to form an image that reflects the meaning of the text~\cite{wikipedia2014calligram}. By blending readability with visual form, calligrams are widely used in logos, advertisements, posters, and book covers to achieve both clarity and aesthetic appeal.
Creating a calligram involves balancing two goals: maintaining legible text and shaping it into a recognizable figure~\cite{maharik2011digitalmicrography,zou2016legible}. Legibility is an important criterion in this setting, because a successful calligram should remain readable while forming a recognizable figure. This dual requirement makes calligram generation a joint optimization problem over textual structure and geometric shape.

Unlike long-text micrography, which typically relies on line-level alignment with minimal glyph deformation~\cite{maharik2011digitalmicrography}, calligrams demand greater geometric control over the placement and deformation of individual glyphs. If the design optimizes solely for shape similarity, letter integrity may be compromised; if it focuses only on readability, the intended figure may become unclear. Thus, the central challenge lies in finding a stable compromise among semantic fidelity, topological coherence, and spatial coverage.

In practice, designers usually choose a target word, find a semantically compatible shape, and then refine letter placement and layout. The dominant cost is typically incurred at the front-end word-to-shape selection stage rather than in downstream local tuning. In other words, the hardest step is identifying structures that are both semantically appropriate and geometrically writable.
\begin{figure}[t]
  \centering
  \includegraphics[width=\columnwidth]{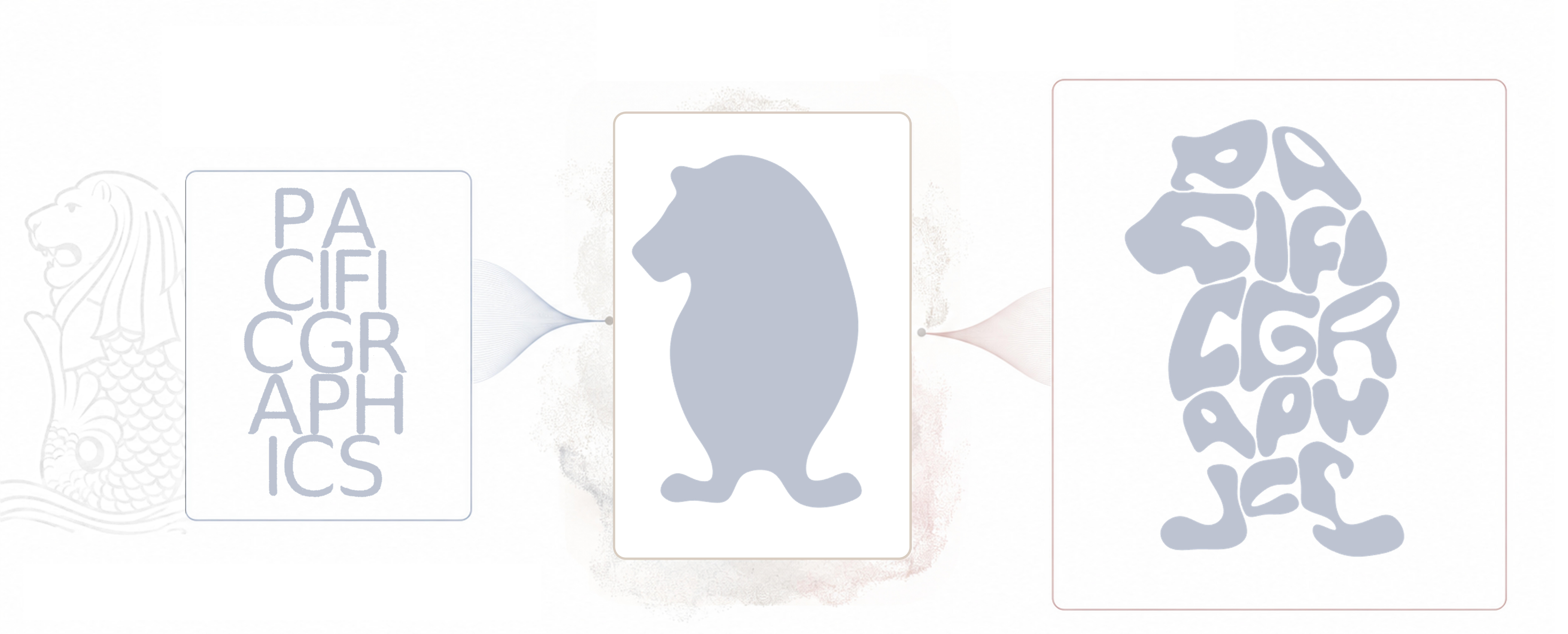}
  \caption{Overview of Telligram. Given the pre-arranged word ``PACIFIC GRAPHICS'' and the semantic prompt ``Merlion'', the method first deforms the glyphs to form a semantic occupancy prior and then generates a readable final calligram result.}
  \label{fig:teaser}
\end{figure}

This practical bottleneck helps explain the boundary of many automatic pipelines. Existing methods have established strong results for given-shape layout, but they typically assume that the target contour is already available. Shape-conditioned approaches such as Calligraphic Packing\cite{xu2007calligraphic} rely on explicit containers and do not address automatic shape acquisition. Legible Compact Calligrams\cite{zou2016legible} improve the readability--fidelity balance, but still depend on manually selected input images and their geometry. Decoupled pipelines \cite{berio2025bsplines} that generate an image first and overlay text afterward often cause letters to concentrate in major body regions and fail to stably cover thin branches and boundary details. Consequently, the expensive front-end stage remains largely unaddressed, and the missing piece is not merely a stronger optimizer, but a unified framework that jointly supports semantic shape formation and letter-level structural writability.

To close this gap, we study \textbf{text-only calligram generation without input images}. The system takes only a target word and a semantic prompt, without a manually provided contour, and directly generates a word-level structure. This shifts the problem from layout inside a fixed container to semantic-to-structure generation under readability constraints.

A second challenge is supervision. There is no dedicated large-scale dataset built around skeleton parameter optimization targets for calligrams. Training an additional learned guidance module in this space is costly, and the valid deformation range of the same letter can change across words, such as the letter \emph{c} in \emph{camel} and \emph{crocodile}. Existing readability signals are often defined at the image level, OCR level, or user-study level, which are useful for evaluation but difficult to convert into stable optimization gradients in the skeleton parameter space. This makes a single learned guidance rule in the skeleton parameter space hard to define and generalize.

Our main view is that, with current training limitations, Variational Score Distillation (VSD)~\cite{wang2023prolificdreamer} is more reliable for predicting occupancy regions (where letters should be placed) than for producing directly readable stroke topology (how strokes connect). Based on this observation, we use a region-first strategy. We first generate a semantic occupancy prior and then convert it into per-letter regions for later glyph filling. This strategy turns a strongly coupled joint problem into a conditional geometric problem, which lowers optimization complexity and reduces instability transfer across stages.

In this paper, we present a training-free text-only pipeline for calligram generation. The method combines semantic occupancy prior formation with later glyph realization and uses hierarchical gradient projection to improve the stability of the optimization stage.

The main contributions of this paper are as follows. First, we formulate text-only calligram generation without an input contour, where semantic shape support must be inferred from text rather than provided in advance. Second, we propose a training-free two-stage framework in which hierarchical gradient projection stabilizes semantic occupancy prior formation under diffusion guidance. Third, we develop a readability-constrained geometric realization stage that converts the occupancy prior into fillable letter regions through lightweight geometric processing.

\section{Related Work}

Recent studies have advanced calligram-related tasks under different assumptions. Some assume a given target shape. Some focus on concept-driven glyph editing. Some study font generation. Others address large-scale text layout. For clarity, we review them by target assumptions, structural scope, and geometric control. This view matches our setting: whole-word calligram generation in an image-free setting.

\subsection{Shape-Conditioned Calligram and Image-Guided Text Composition}

Shape-conditioned methods are the most established direction in automatic calligram generation. Their common assumption is that the target contour is known. Calligraphic Packing \cite{xu2007calligraphic} divides a container region and assigns letters to local subregions, giving strong contour-filling results. This design is effective because its objectives are explicit and its constraints are geometric. Its limitation is equally clear: it solves layout \emph{inside} a given shape, but does not generate that shape from semantic input.

Legible Compact Calligrams \cite{zou2016legible} improve readability by combining contour skeletons, local protrusion handling, and letter-anchor matching in a hierarchical pipeline. A key lesson is that readability should be optimized together with geometric fitting, not only in post-processing. However, the method still needs external target geometry, so it cannot decide whole-word structure on its own in an image-free setting.

Related image-guided typography and composition methods, including ornamental typeface synthesis \cite{zhang2017ornamental}, word paintings \cite{zhang2022wordpaintings}, and ShapeWordle \cite{wang2020shapewordle}, also show that text can serve as a strong visual primitive. In most cases, global structure is controlled by an external image or shape, while text is aligned and filled locally. These works offer useful geometric operations, but they do not directly solve image-free semantic structure generation. In short, shape-conditioned calligram layout is mature, but it still assumes given target geometry.

\subsection{Semantic Typography and Word-as-Image}

With diffusion models and text-image prior knowledge, recent work has shifted from fitting predefined contours to making text itself carry semantic meaning. Word-as-Image \cite{iluz2023wordasimage} is a representative example. It injects concept cues into letterforms, enabling local correspondence between glyph shape and concept. Its key contribution is to reduce dependence on external contours and show that semantic-driven glyph editing is feasible.

At the task level, these methods mainly target single letters or short character groups. This is closer to concept-driven glyph editing than to full word-shape construction. Put simply, they show how one character can reflect a concept, but not how a full word can become a coherent figurative object with readable order.

Later diffusion-based artistic typography methods, such as DS-Fusion \cite{tanveer2023dsfusion} and Anything to Glyph \cite{wang2023anythingtoglyph}, improve style richness and material appearance, but still focus on pixel-space stylization and often rely on extra shape cues or layout priors. More general multimodal generative models, such as GPT Image 2.0 \cite{openai2026gptimage2}, show a similar limit: they can usually preserve either global contour or local letter readability, but not both. Typical outputs are semantic images built from repeated copies of one word, or non-semantic structures formed by only a few letters. Thus, a gap remains for strict whole-word calligram generation.

Khattat \cite{hussein2024khattat} combines semantic prompting, font selection, and text-recognition readability constraints in one framework, marking an important step for multi-character semantic typography. Related work such as FontCLIP \cite{tatsukawa2024fontclip} supports this direction. The strength of this line is joint optimization over semantics, style, and legibility. Still, it targets semantic stylization and local readability improvement rather than full figurative whole-word structure in an image-free setting. For our task, it offers useful supervision, but not a direct structural solver.

\subsection{Font Generation, Glyph Modeling, and Layout Synthesis}

Font and glyph generation methods mainly study how to produce high-quality, editable glyphs. Their main goal is not to induce word-level figurative structure. DeepVecFont \cite{wang2021deepvecfont}, VecFontSDF \cite{xia2023vecfontsdf}, and VecFusion \cite{thamizharasan2024vecfusion} improve vector glyph quality through joint image-vector representation, signed-distance-shape modeling, and diffusion generation. These methods provide strong foundations for representation quality, style consistency, and editability.

Their typical scope remains in font space, such as reconstruction, transfer, and stylization, rather than whole-word object formation. High local glyph quality does not automatically produce valid word-level structure. This suggests that local glyph modeling and word-level structural optimization should be treated as related but separate layers.

Related work on text layout synthesis, such as text-logo layout \cite{wang2022textlogo}, focuses on character scale, affine transforms, line breaking, and global composition balance. It usefully distinguishes local glyph quality from word-level organization, which is directly relevant to whole-word structure.

\subsection{Large-Scale Text Layout and Word Clouds}

Word-cloud and large-scale text layout research has long studied the placement of many text units within bounded regions. Core concerns include collision avoidance, density control, and readability balance. Classical work \cite{kaser2007tagcloud} analyzes 2D layout cost and space use. Later studies add temporal stability via context-preserving dynamic word clouds \cite{cui2010contextwordcloud}, interaction constraints via ManiWordle \cite{koh2010maniwordle}, and multi-attribute organization via FacetClouds \cite{waldner2013facetclouds}.
This line offers reusable layout tools, but also shows that layout is already strongly constrained even when shape is known.

Word-cloud settings are fundamentally different from calligrams. Word clouds usually treat words or short phrases as independent units and mainly optimize position, scale, and orientation. Whole-word calligrams require letter-level continuity and preserve reading order. Their shape constraints are also usually external, for example, in shape-bounded ShapeWordle settings \cite{wang2020shapewordle}. Thus, word-cloud methods are better treated as layout toolkits than as direct solutions for image-free calligram generation.

\section{Prior Knowledge}

\subsection{VSD Semantic Prior}

Score Distillation Sampling (SDS) uses a pretrained diffusion model as a prior and performs conditional generation by differentiable parameter optimization. The main idea is to avoid training a new generator. Instead, SDS uses denoising score signals from the diffusion model to provide semantic gradient constraints on the target parameters. Variational Score Distillation (VSD) can be viewed as a variational extension of SDS. SDS treats the optimization target as a single deterministic point, while VSD lifts it to a random variable over a distribution and improves gradient stability and diversity through a distribution-level variational view~\cite{poole2022dreamfusion,wang2023prolificdreamer}.

Let the optimizable parameters be $\Theta$, and let $x(\Theta)$ be the observation produced by parameterized rendering. At noise timestep $t$, a common SDS gradient form is
\begin{equation}
\nabla_{\Theta}\mathcal{L}_{\mathrm{SDS}}
\propto
\mathrm{E}_{t,\epsilon}\left[w(t)\left(\hat{\epsilon}_{\phi}(x_t,y,t)-\epsilon\right)
\frac{\partial x}{\partial \Theta}\right],
\end{equation}
where $x_t$ is the noisy observation at timestep $t$, $\epsilon$ is the sampled Gaussian noise, $y$ is the text condition, and $\hat{\epsilon}_{\phi}$ is predicted by the frozen diffusion model. On top of this, VSD extends point-estimate optimization to distribution-level variational matching. This helps reduce oversaturation, premature convergence, and instability that can appear in single-point SDS optimization.

In our method, VSD provides semantic guidance for the current glyph structure. Specifically, the glyph parameters $\Theta$ are rendered into an occupancy image $x(\Theta)$, and the diffusion prior then produces semantic gradients $\nabla_{\Theta}\mathcal{L}_{\mathrm{VSD}}$ that guide the current structure toward the target semantic concept.

Overall, VSD supplies semantic gradients, while geometric parameterization keeps the optimization structured and controllable.

\subsection{Voronoi Partition}

Voronoi partition is a standard geometric tool for region assignment and boundary construction. Given a set of sample points $\{\mathbf s_k\}_{k=1}^K$ in the plane, the Voronoi cell of the $k$-th point is defined as
\begin{equation}
V_k=\{\mathbf x\in\mathbb R^2 \mid d(\mathbf x,\mathbf s_k)\le d(\mathbf x,\mathbf s_j),\ \forall j\neq k\},
\end{equation}
where $\mathbb R^2$ is the 2D plane and $d(\mathbf x,\mathbf s_k)$ is the Euclidean distance between points $\mathbf x$ and $\mathbf s_k$. In other words, $V_k$ contains all points that are closer to $\mathbf s_k$ than to any other point.
This construction divides space into non-overlapping regions according to nearest-site distance.

In graphics and geometry processing, Voronoi partition is often used to separate neighboring structures, assign local support regions, and construct boundary-aware constraints. When the sites are sampled from skeletons or centerlines, the induced partition gives a natural way to decide which spatial area belongs to which structure. The boundaries between Voronoi cells can also be interpreted as separating lines that preserve local detail and reduce unwanted merging between nearby components. In our method, Voronoi partition is used as a boundary-aware spatial separator during later geometric realization.

\begin{figure*}[t]
  \centering
  \includegraphics[width=\textwidth]{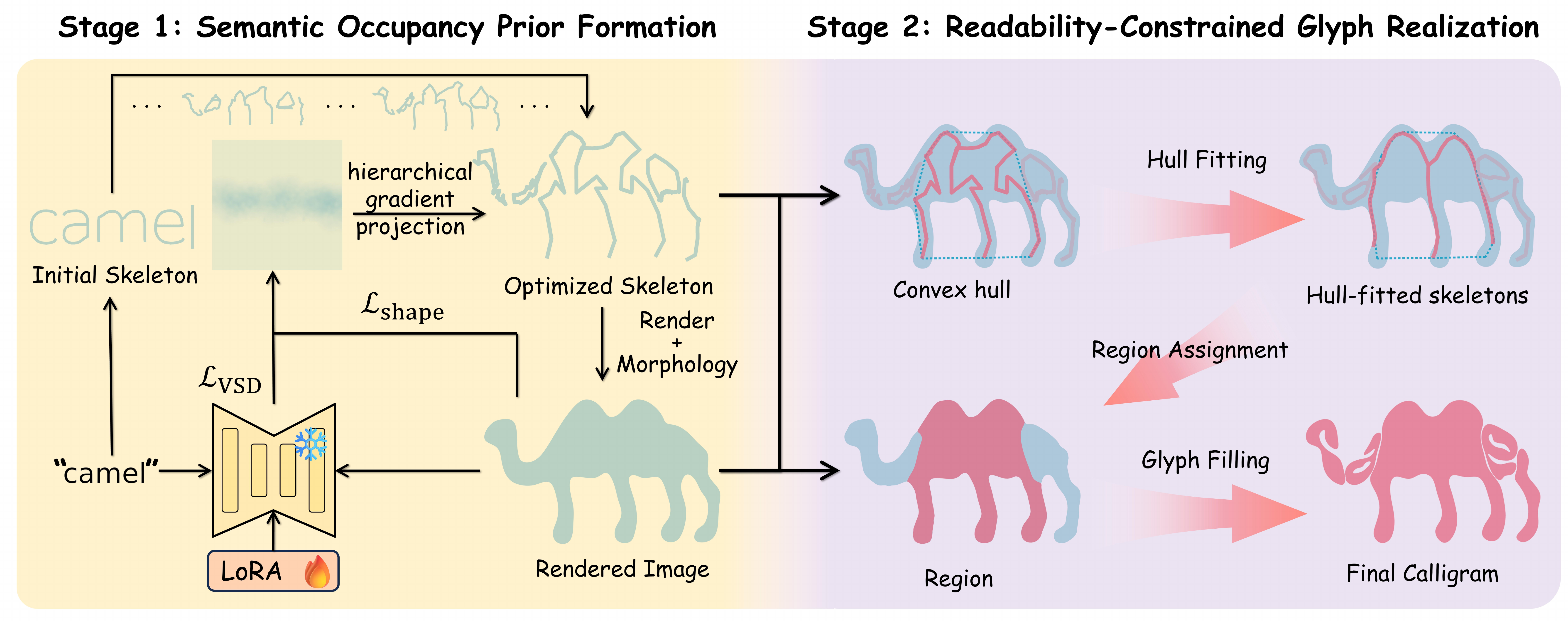}
  \caption{Overview of our two-stage method. In Stage 1, the input word is converted into parameterized letters and optimized under VSD semantic guidance, shape regularization, and hierarchical gradient projection to produce a semantic occupancy prior. In Stage 2, the resulting occupancy mask and skeleton are converted into a readable final calligram through convex-hull extraction, hull-guided skeleton fitting, region assignment, and glyph filling.}
  \label{fig:pipeline}
\end{figure*}

\section{Method}

This method is a training-free, two-stage pipeline for text-only calligram generation. It aims to form a figurative word-level structure while keeping letter readability and geometric feasibility. The first stage, Semantic Occupancy Prior Formation, builds a semantic occupancy prior from parameterized letters under diffusion guidance and lightweight shape regularization. The second stage, Readability-Constrained Glyph Realization, converts this occupancy prior into per-letter regions and readable glyphs through lightweight geometric processing, including convex-hull extraction, hull-guided skeleton fitting, region assignment, and Voronoi-assisted filling. As shown in Fig.~\ref{fig:pipeline}, the two stages separate semantic structure formation from topology-sensitive glyph realization.

Skeleton-based parameterization provides a compact deformation space that preserves letter topology more reliably under large semantic motion. In our setting, Bezier-based control-point deformation was more sensitive to hyperparameter choice and more likely to damage recognizable letter structure. The skeleton representation therefore provides a more stable balance between semantic mobility and topological stability. A two-stage design is used because semantic formation and readability preservation act on competing geometric tendencies. Stage 1 first forms a semantically plausible occupancy prior, and Stage 2 then restores letter structure under that support. Directly coupling these two goals in a single optimization loop was less stable in our experiments, so the final pipeline uses a single semantic-then-realization pass.

\subsection{Semantic Occupancy Prior Formation}

This stage takes a word and a semantic prompt and optimizes the letters toward a recognizable semantic occupancy prior. The goal is not to directly produce readable final glyphs, but to obtain a stable occupancy prior that captures where letters should be placed.

We first convert the input word into parameterized letters and initialize an optimizable state $\Theta=\{\mathbf T,\mathbf S,\mathbf R,\mathbf J\}$, where $\mathbf T$, $\mathbf S$, and $\mathbf R$ denote per-letter translation, scale, and rotation, and $\mathbf J$ denotes local skeleton coordinates. Each glyph is rendered from a skeleton with fixed radii, where each skeleton segment is given a fixed stroke width, and its occupied area is approximated by a union of capsule primitives, that is, short rounded stroke pieces.

From this state, differentiable rendering produces a soft mask $\alpha\in[0,1]^{H\times W}$, where $H\times W$ is the image size, and each pixel value indicates how strongly that pixel is covered. We then apply shape-growing and shape-shrinking operations to obtain a more semantically usable mask. Specifically, we use LogSumExp-based soft pooling, a smooth approximation of max/min pooling, to define a soft close operation, which first expands and then shrinks the mask,
\begin{equation}
\operatorname{Close}_r^{\tau}(\alpha)=\varepsilon_r^{\tau}(\delta_r^{\tau}(\alpha)).
\end{equation}
Here $\delta_r^{\tau}$ and $\varepsilon_r^{\tau}$ denote soft dilation and soft erosion with radius $r$ and temperature $\tau$, respectively. Dilation means expanding the mask outward, erosion means shrinking it inward, $r$ sets the operation range, and $\tau$ controls how sharp or smooth the approximation is. The role of the soft morphological close is to introduce weak pseudo-connections in local gaps, so semantic guidance does not need to force true letter overlap in order to form a coherent figure.

To avoid false boundary connections, we use a boundary-band mask $B\in\{0,1\}^{H\times W}$, where 1 marks a narrow image-border band in which closing is suppressed, and apply a border-safe close. We then run reconstruction-style hole filling from boundary markers:
\begin{equation}
m^{(t+1)}=\min\big(\delta_r(m^{(t)}),\beta\big),
\qquad
\beta=1-\alpha_{\mathrm{safe}},
\end{equation}
and obtain $\alpha_{\mathrm{fill}}=1-m^{(*)}$ after convergence, where $\alpha_{\mathrm{safe}}$ is the border-safe mask after suppressing closing near the image boundary. Here, $\delta_r$ denotes standard binary dilation in the reconstruction step, that is, one step of binary mask expansion with radius $r$. $\beta$ is the background area that filling is allowed to enter, and $m^{(*)}$ is the final marker after the iteration stops.

The optimized objective is written as
\begin{equation}
\mathcal L_{\mathrm{total}}=w_{\mathrm{sem}}\mathcal L_{\mathrm{VSD}}+w_{\mathrm{shape}}\mathcal L_{\mathrm{shape}}.
\end{equation}
Here, $\mathcal L_{\mathrm{VSD}}$ is the semantic guidance from the diffusion model, applied to the rendered and morphologically processed mask. On the rendered mask, we use two lightweight geometric terms. Let $\Omega$ denote the full pixel grid of the rendered occupancy image. Let $\alpha_k(\mathbf x)$ be the rendered body mask value of the $k$-th letter at pixel $\mathbf x$, and let $\alpha(\mathbf x)$ denote the full rendered mask. We define
\begin{equation}
\mathcal L_{\mathrm{overlap}}=
\frac{1}{|\Omega|}
\sum_{\mathbf x\in\Omega}
\operatorname{ReLU}\!\left(\sum_{k=1}^{K}\alpha_k(\mathbf x)-1\right),
\end{equation}
\begin{equation}
\mathcal L_{\mathrm{single}}=
\frac{1}{|\Omega|}
\sum_{\mathbf x\in\Omega}
\alpha(\mathbf x)\bigl(1-c(\mathbf x)\bigr),
\end{equation}
where $c(\mathbf x)$ is a soft connectivity value obtained by iterative masked dilation from seed regions, i.e., repeated masked expansion from initial seed pixels; it measures whether pixel $\mathbf x$ belongs to the connected main component. $\mathcal L_{\mathrm{overlap}}$ is computed on the rendered letters and discourages excessive inter-letter interference. $\mathcal L_{\mathrm{single}}$ encourages all letters to contribute to the main semantic body, 
preventing VSD from converging to a degenerate solution
in which one enlarged letter explains most of the shape while the remaining letters shrink into small isolated blobs. The shape regularizer is then defined as
\begin{equation}
\mathcal L_{\mathrm{shape}}=\lambda_{\mathrm{overlap}}\mathcal L_{\mathrm{overlap}}+\lambda_{\mathrm{single}}\mathcal L_{\mathrm{single}}.
\end{equation}
Together, these terms keep the semantic optimization from collapsing into unreadable or heavily unbalanced letter layouts.

\begin{figure}[t]
  \centering
  \includegraphics[width=\columnwidth]{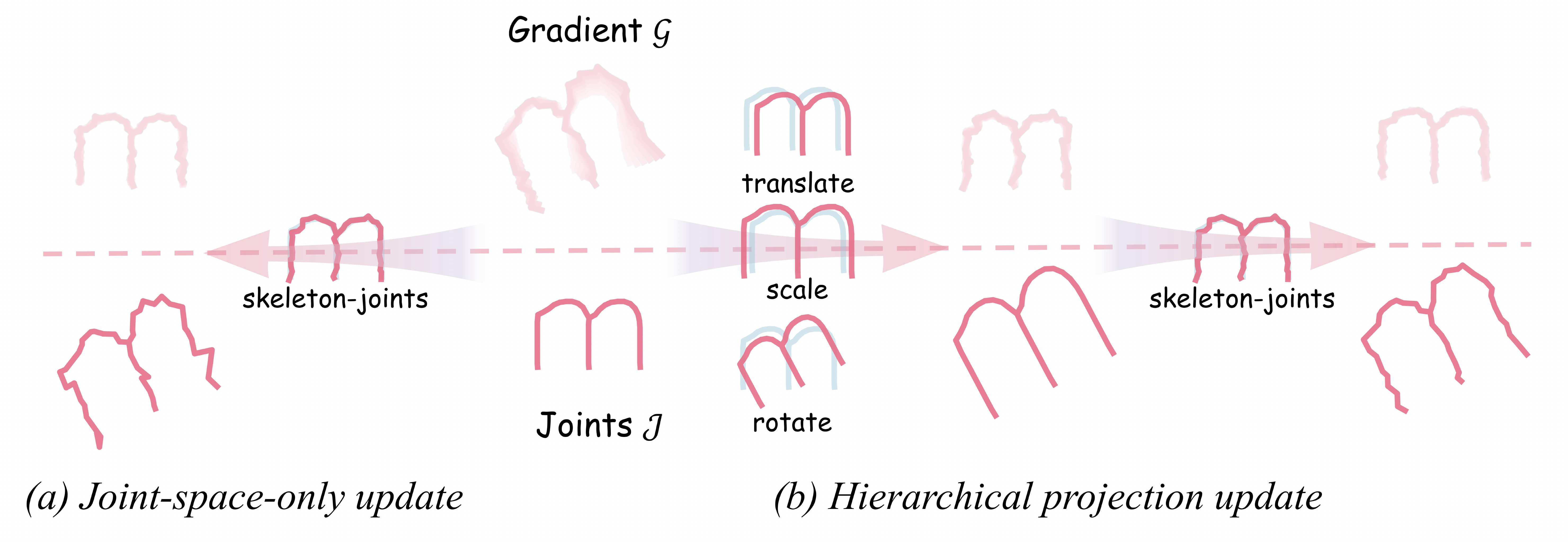}
  \caption{Illustration of joint-space-only update and hierarchical projection on the skeleton of the letter m. The pale blurry shape visualizes the current gradient-induced motion field, and the pink curve shows the current skeleton. In (a), the full gradient is directly injected into joint space, so coarse displacement and local deformation remain entangled throughout the update. In (b), the update flow is split into stages: low-dimensional affine motion first absorbs coarse semantic displacement, and only the residual local signal is passed to the skeleton variables. This ordering makes the optimization trajectory easier to follow and reduces unstable shortcuts dominated by coarse deformation.}
  \label{fig:hierarchical_projection}
\end{figure}

The remaining question is how to update these variables without collapsing the layout. A common strategy in packing or collage optimization is to impose explicit lower and upper bounds on per-element scale ~\cite{shao2026score}, so that no element becomes too large or too small. However, this strategy is too restrictive for compact calligrams. To form a semantic figure, some letters may need to become noticeably larger than others, such as the letter \emph{m} in a camel-shaped word. Hard scale bounds can therefore block useful semantic deformation. Without such control, optimization over coarse global deformation and fine local deformation often converges to degenerate shortcuts, where a few letters dominate the main body, and the others nearly disappear. We therefore use a hierarchical gradient projection as a soft alternative to hard scale bounds. As illustrated in Fig.~\ref{fig:hierarchical_projection}, the update is organized in a coarse-to-fine order instead of directly injecting the full gradient into skeleton joints.

This design avoids two unstable extremes. A manually staged schedule for low-dimensional affine deformation and high-dimensional skeleton deformation is hard to tune. Directly injecting the full gradient into skeleton joints makes the update sensitive to high-frequency noise, while smoothing such joint-space updates may also suppress useful coarse semantic motion. We therefore let low-dimensional affine deformation absorb the large structured part of the gradient first, and pass only the remaining signal to the local skeleton variables. The subsequent smoothing step is then applied only to this residual local update, which reduces information loss and improves stability. Radius freezing, meaning that stroke width is kept fixed, is maintained throughout this stage.

Let $\mathbf G=-\nabla_{\mathbf J}\mathcal L_{\mathrm{total}}$ be the back-propagated force field on skeleton joints. We first project this gradient onto a low-dimensional affine subspace that includes translation, isotropic scaling, rotation, and anisotropic scaling,
\begin{equation}
\begin{aligned}
\mathbf a_{\mathrm{aff}}
&=\arg\min_{\mathbf a}
\|\mathbf G-\mathbf B_{\mathrm{aff}}\mathbf a\|_{2}^{2}
+\lambda_{\mathrm{aff}}\|\mathbf a\|_{2}^{2}, \\
\mathbf G^{(\mathrm{loc})}
&=\mathbf G-\mathbf B_{\mathrm{aff}}\mathbf a_{\mathrm{aff}} .
\end{aligned}
\end{equation}
where $\mathbf B_{\mathrm{aff}}$ denotes the basis matrix of the low-dimensional affine deformation subspace, meaning the set of allowed coarse deformation directions, and $\mathbf a_{\mathrm{aff}}$ contains the corresponding affine coefficients. We then decompose the remaining signal into a graph-smooth skeleton component and a high-frequency residual on the skeleton graph $\mathcal T=(\mathbf J,\mathbf E)$, where $\mathbf E$ is the set of skeleton edges connecting neighboring joints.
\begin{equation}
\begin{aligned}
\mathbf G_{\mathrm{skel}}
&=(\mathbf I+\lambda_{\mathrm{skel}}\mathbf L)^{-1}
\mathbf G^{(\mathrm{loc})}, \\
\mathbf G_{\mathrm{res}}
&=\mathbf G^{(\mathrm{loc})}-\mathbf G_{\mathrm{skel}} .
\end{aligned}
\end{equation}
where $\mathbf L$ is the graph Laplacian of $\mathcal T$, a standard matrix that encodes which skeleton joints are neighbors and therefore favors smooth changes along the skeleton. The residual term is computed only to make the decomposition explicit, and is discarded in the final update to avoid unstable high-frequency local distortion. The final update is written as
\begin{equation}
\Delta\Theta_{\mathrm{aff}}=-\eta_{\mathrm{aff}}\mathbf a_{\mathrm{aff}},
\end{equation}
\begin{equation}
\Delta\mathbf J=-\eta_{\mathrm{skel}}\mathbf G_{\mathrm{skel}},
\end{equation}
where $\Delta\Theta_{\mathrm{aff}}$ updates the affine variables in $\Theta$ and $\Delta\mathbf J$ updates the skeleton coordinates, that is, the point locations on the letter centerline. Here $\eta_{\mathrm{aff}}$ and $\eta_{\mathrm{skel}}$ are the step sizes for the affine and skeleton updates. The full state update is therefore
\begin{equation}
\Theta_{t+1}=\{\mathbf T_{t+1},\mathbf S_{t+1},\mathbf R_{t+1},\mathbf J_{t+1}\}.
\end{equation}
This hierarchical projection reduces tuning burden while preserving the freedom needed for semantic deformation. In practice, large semantic motion is first absorbed by low-dimensional affine transforms, while local skeleton deformation is used only when necessary. The output of Stage 1 is therefore a semantic occupancy prior rather than a directly readable final glyph layout.

\subsection{Readability-Constrained Glyph Realization}

The output of Stage 1 provides a semantic occupancy prior, but it is not yet a readable final calligram. This stage is a lightweight geometric post-processing step that converts the Stage 1 occupancy mask and skeleton into per-letter regions and readable letter shapes. The process follows four steps: convex-hull extraction, hull-guided skeleton fitting, region assignment, and Voronoi-assisted filling. Since this stage is not the main contribution, we keep it lightweight and training-free.

We begin by extracting a convex hull, the smallest convex region that covers the input skeleton support, of each letter. This hull provides a coarse outer shape that preserves how far the letter can extend inside the occupancy mask. We then fit the input skeleton into this hull to obtain a more readable skeleton configuration. The purpose of this step is not to exactly recover the original glyph, but to restore a clearer letter structure before final filling.

Let the convex hull of the $k$-th letter be $\Pi_k$ and let $\mathcal S_k$ denote its fitted skeleton, that is, the centerline used to rebuild that letter. The fitted skeleton is constrained to remain geometrically compatible with $\Pi_k$, so that it follows the allocated semantic mass while avoiding excessive distortion.

Once fitted skeletons are available, we assign a region to each letter. The per-letter region is determined jointly by the Stage 1 occupancy mask and the fitted skeleton layout. Intuitively, the mask provides the available semantic support, while the fitted skeleton determines which part of that support should belong to each letter.

Formally, let $M\subset\mathbb Z^2$ be the Stage 1 occupancy mask, where $\mathbb Z^2$ denotes the 2D integer pixel grid. For each pixel $\mathbf x\in M$, we assign it to the nearest fitted skeleton, where $d(\mathbf x,\mathcal S_k)$ is the Euclidean distance from pixel $\mathbf x$ to skeleton $\mathcal S_k$,
\begin{equation}
L(\mathbf x)=\arg\min_k d(\mathbf x,\mathcal S_k),
\end{equation}
and obtain per-letter regions $M_k=\{\mathbf x\in M\mid L(\mathbf x)=k\}$. These per-letter regions serve as the support for final glyph filling.

With a fitted skeleton $\mathcal S_k$ and its region $M_k$, the remaining problem is to decide how strongly the skeleton should be thickened so that the letter fills the region while remaining readable. We therefore generate candidate fillings by skeleton-radius dilation, meaning that we thicken the skeleton by a radius $r$,
\begin{equation}
G_k(r)=\operatorname{Dilate}(\mathcal S_k,r)\cap M_k,
\end{equation}
and use TrOCR~\cite{li2021trocr}, a text recognition model, as an auxiliary scorer to select a radius that balances readability and region coverage. Here, $\operatorname{Dilate}(\mathcal S_k,r)$ means the set of pixels whose distance to skeleton $\mathcal S_k$ is at most $r$.

Readability score alone is not sufficient. For some letters, a very small radius keeps the skeleton recognizable and therefore receives a relatively high score, but leaves the region underfilled. Conversely, a large radius may fill the region well but erase thin structures and lower readability. To reduce this failure mode, we introduce a Voronoi-based constraint for letters whose local details are easily removed by direct thickening.

Specifically, we compute a Voronoi partition from the fitted skeletons, meaning that each pixel is assigned to its nearest skeleton, and use the induced boundary lines as local separators. After thickening these boundary lines into a narrow protection band, meaning a thin forbidden strip near letter boundaries, we subtract the band from the target region before final filling. Denoting this protection band by $C_k$, the final realization is written as
\begin{equation}
\hat G_k=(\operatorname{Dilate}(\mathcal S_k,r_k)\cap M_k)\setminus C_k,
\end{equation}
where $r_k$ is the selected filling radius for the $k$-th letter, that is, the final stroke thickness used for that letter. This simple refinement preserves local details and inter-letter separation, while still allowing each glyph to fill its assigned region as much as possible.

In summary, the method consists of two parts. Semantic Occupancy Prior Formation generates stable occupancy through rendered-mask semantic supervision, lightweight shape regularization, and hierarchical gradient projection, while keeping stroke width fixed during optimization. Readability-Constrained Glyph Realization converts this occupancy prior into readable glyphs through convex-hull extraction, hull-guided skeleton fitting, region assignment, and Voronoi-assisted filling. This design separates semantic occupancy formation from readable glyph realization, while keeping the second stage lightweight and geometric.

\section{Experiments}

This section evaluates both the optimization behavior of the proposed framework and the quality of the generated calligrams. We first study how the proportion of early low-dimensional affine updates affects optimization behavior, where low-dimensional affine updates refer to coarse global motion such as translation, rotation, and scaling. We then test whether hierarchical gradient projection improves semantic alignment while avoiding collapse patterns caused by overly strong coarse deformation. Finally, we report OCR-based readability results and representative qualitative ablations. Unless otherwise noted, all table values in this section are reported in units of $\times 10^{-2}$.

\begin{table}[t]
\centering
\small
\caption{Runtime and platform settings used in the experiments.}
\label{tab:runtime_details}
{\setlength{\tabcolsep}{5pt}
\renewcommand{\arraystretch}{1.08}
\begin{tabular}{p{0.34\columnwidth}p{0.50\columnwidth}}
\toprule
Item & Value \\
\midrule
Stage 1 steps & 2000 \\
Semantic resolution & $96\times96$ \\
VSD model & DeepFloyd-IF Stage I ~\cite{deepfloyd2023if} \\
GPU & NVIDIA Quadro P6000 \\
GPU memory & 24 GB \\
Stage 1 runtime & about 6 min per letter \\
Stage 2 runtime & within 5 min \\
\bottomrule
\end{tabular}}
\end{table}

\begin{table}[t]
\centering
\small
\caption{Core hyperparameters used in the experiments.}
\label{tab:key_hparams}
{\setlength{\tabcolsep}{5pt}
\renewcommand{\arraystretch}{1.08}
\begin{tabular}{ll}
\toprule
Parameter & Value \\
\midrule
Soft-close temperature $\tau$ & 0.6 \\
Word-level close radius $r$ & 15 \\
Semantic weight $w_{\mathrm{sem}}$ & 120 \\
Single-component weight $\lambda_{\mathrm{single}}$ & 12 \\
Overlap weight $\lambda_{\mathrm{overlap}}$ & 1 \\
Smoothing weight $\lambda$ & 0.5 \\
\bottomrule
\end{tabular}}
\end{table}

For quantitative evaluation, we use OpenAI CLIP ViT-B/32~\cite{radford2021clip}. CLIPScore is averaged over 10 different prompts and 15 different random seeds for each compared setting. We compare three update rules. \textbf{Direct joint update} applies the full gradient directly to the skeleton joints. \textbf{Smoothed joint update} first computes a graph-smoothed joint update \(\mathbf G_{\mathrm{smooth}}=(\mathbf I+\lambda \mathbf L)^{-1}\mathbf G\) and then applies it to the joints, where \(\mathbf L\) is the skeleton graph Laplacian and \(\lambda\) is the smoothing weight. \textbf{Hierarchical update} first lets low-dimensional affine deformation absorb the structured part of the gradient, and then applies the same smoothed joint update only to the remaining local signal.

\subsection{Coarse-Deformation Ratio Analysis}

We study how the ratio of early low-dimensional affine updates affects optimization behavior. Table~\ref{tab:lowdim_ratio} summarizes the four schedule settings and their quantitative results. We use CLIPScore as the primary metric. We also report \emph{max region ratio} and \emph{region variance} only as descriptive indicators of how concentrated or imbalanced the final letter areas become under different schedules.

\begin{table}[t]
\centering
\small
\caption{Effect of the early low-dimensional affine update ratio. 
None corresponds to 0\% early affine steps, while Low, Medium, 
and High are stochastic schedules centered around approximately 
25\%, 50\%, and 75\% of the optimization process. 
All values are reported in units of $\times 10^{-2}$.}
\label{tab:lowdim_ratio}

{\setlength{\tabcolsep}{4pt}
\renewcommand{\arraystretch}{1.15}

\begin{tabular}{p{0.24\columnwidth}ccc}
\toprule
\shortstack[l]{Early low-dim.\\update ratio}
& CLIPScore $\uparrow$
& Max ratio
& Reg. var. \\
\midrule
High   & 33.314 & 42.446 & 1.735 \\
Medium & 33.566 & 39.232 & 1.285 \\
Low    & \underline{33.868} & 33.178 & 0.994 \\
None   & \textbf{34.227} & 32.530 & 0.937 \\
\bottomrule
\end{tabular}
}
\end{table}

The main trend is reflected by CLIPScore. Under the staged affine schedule, earlier participation of high-dimensional skeleton deformation consistently leads to better semantic alignment. This suggests that relying too much on early low-dimensional affine motion encourages coarse shortcuts, while earlier skeleton deformation gives the optimization more freedom to match the target concept. The max region ratio and region variance provide only a descriptive view of how concentrated the letter areas become under different schedules. They are not used here as direct quality criteria. Taken together, these results show that the main difficulty is not simply whether coarse motion is enabled, but how much of the optimization budget is assigned to it at early stages. This also explains why manually staged schedules are difficult to tune in practice.

\subsection{Hierarchical Projection Analysis}

We compare the three update rules defined above under the same overall budget: \textbf{Direct joint update}, \textbf{Smoothed joint update}, and \textbf{Hierarchical update}. The first is equivalent to the \textbf{None} setting in Table~\ref{tab:lowdim_ratio}; the two names describe the same method from two different views. CLIPScore remains the primary metric, while max ratio and region variance are used only to describe size concentration and area imbalance.

\begin{table}[t]
\centering
\small
\caption{Main comparison of optimization strategies. The full method is expected to provide the best semantic alignment while avoiding dominant-letter collapse. All values are reported in units of $\times 10^{-2}$.}
\label{tab:main_compare}
{\setlength{\tabcolsep}{4pt}
\renewcommand{\arraystretch}{1.08}
\begin{tabular}{p{0.30\columnwidth}ccc}
\toprule
Method & CLIPScore $\uparrow$ & Max ratio & Reg. var. \\
\midrule
Direct joint update & 34.227 & 32.530 & 0.937 \\
Smoothed joint update & \underline{34.288} & 31.581 & 0.742 \\
Hierarchical update & \textbf{34.314} & 36.845 & 1.593 \\
\bottomrule
\end{tabular}}
\end{table}

The hierarchical update achieves the highest CLIPScore among the compared settings. This result indicates that explicitly separating low-dimensional affine motion from high-dimensional skeleton deformation improves semantic alignment more effectively than either direct joint update or smoothed joint update alone. Notably, the hierarchical update does not aim to minimize size imbalance. Its max ratio and region variance can be higher than those of direct joint update, because it allows more flexible letter-size adaptation. The advantage is that this flexibility is introduced in a structured way, so semantic alignment improves without relying on unstable joint-wise noise.

Overall, the experiments support two conclusions: high early low-dimensional update ratios favor dominant-letter shortcuts, while the hierarchical update provides the best semantic alignment with structured and flexible letter-size adaptation.

\begin{figure*}[t]
  \centering
  \includegraphics[width=\textwidth]{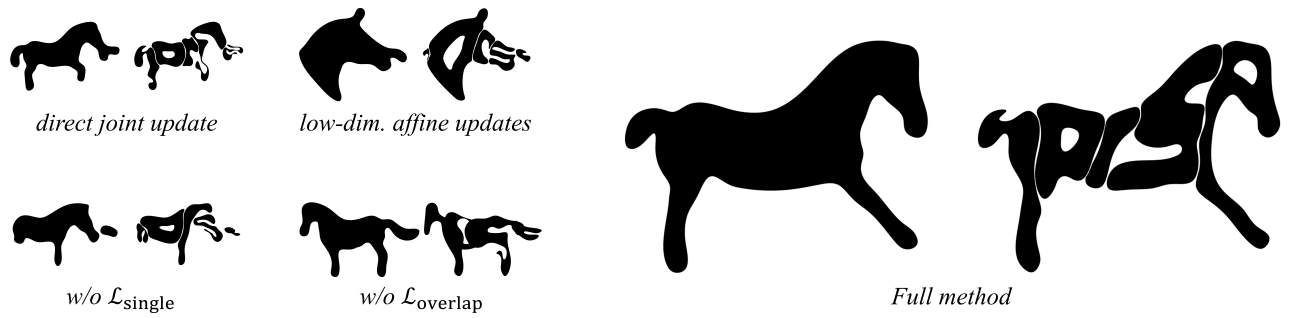}
  \caption{Representative qualitative ablations for Stage 1 modules. From left to right, we show the effects of prolonged early affine updates, skeleton-only deformation, removing $L_{\mathrm{single}}$, and removing $L_{\mathrm{overlap}}$. These examples illustrate how the individual components affect dominant-letter collapse, semantic motion, disconnected small letters, and excessive inter-letter interference.}
  \label{fig:main_ablation}
\end{figure*}

\subsection{OCR-Based Readability Evaluation}

Since calligram quality depends on both semantic shape and character recognizability, we complement the semantic evaluation with an OCR-based readability experiment on cropped single-character images. Although TrOCR is used in the local candidate-selection step of Stage 2, the readability evaluation here is conducted with separate OCR backbones so that the final analysis remains independent of that selection process. Because the inputs are already character-level crops, we use character recognizers rather than full text-detection pipelines. We report results from two representative backbones, CRNN~\cite{shi2015crnn} and PARSeq~\cite{bautista2022parseq}, which provide two recognition views on the generated glyphs.

For each generated word, we split the final calligram into individual letter images and evaluate whether each crop is still recognized as its ground-truth character. We report character-level top-1 accuracy for both recognizers and, for CRNN, also the mean ground-truth posterior $P(\mathrm{gt}\mid x)$. OCR is used here as an auxiliary readability measure rather than a complete perceptual metric, since our method also explicitly targets smooth boundaries and the suppression of spurious branches.

\begin{table}[t]
\centering
\small
\caption{OCR-based readability evaluation on cropped single-character images. PARSeq and CRNN provide two independent recognition views. For CRNN, we also report the mean ground-truth posterior. ``Berio et al.'' and ``Zou et al.'' denote results reproduced from~\cite{berio2025bsplines,zou2016legible}, while ``Fleming'' denotes the expert-designed examples shown in Fig.~\ref{fig:comparisons}.}
\label{tab:ocr_eval}
\setlength{\tabcolsep}{4pt}
\renewcommand{\arraystretch}{1.08}
\begin{tabular}{p{0.19\columnwidth}ccc}
\toprule
Group & PARSeq Acc. $\uparrow$ & CRNN Acc. $\uparrow$ & Mean $P_{\mathrm{gt}}$ $\uparrow$ \\
\midrule
Berio et al. & 38.89 & 22.22 & 0.135 \\
Fleming      & 46.00 & 50.00 & 0.244 \\
Zou et al.   & 56.00 & 58.00 & 0.286 \\
Ours         & 44.00 & 38.00 & 0.220 \\
\bottomrule
\end{tabular}
\end{table}

As shown in Table~\ref{tab:ocr_eval}, the overall trend is consistent across both recognizers. Zou et al. achieve the highest OCR accuracy, while Berio et al. give the lowest OCR scores among the compared methods. Ours performs better than Berio et al. under the same no-mask setting and remains comparable to the expert-designed Fleming examples on PARSeq and on the CRNN mean ground-truth posterior.

Boundary style may partly explain the remaining gap. The Fleming examples are manually designed, and our method uses smooth geometric realization. In contrast, Zou et al. produce sharper contours that contain local discriminative cues easily recognized by OCR models.
Accordingly, the OCR results should be interpreted together with boundary quality and branch suppression rather than in isolation. CRNN top-1 accuracy, in particular, is lower for smooth-boundary results than for sharper-contour baselines.

\subsection{Qualitative Ablation Study}

Fig. ~\ref{fig:main_ablation} provides representative qualitative ablations for the Stage 1 modules. Prolonged early affine updates tend to produce a dominant-letter shortcut, while skeleton-only deformation weakens semantic alignment. Removing $L_{\mathrm{single}}$ allows detached small components, and removing $L_{\mathrm{overlap}}$ increases inter-letter interference. Taken together, these ablations show that the full Stage 1 design balances coarse semantic motion, local structural adaptation, whole-word connectivity, and inter-letter separation.

\begin{figure*}[p]
  \centering
  \includegraphics[width=\textwidth,height=0.92\textheight,keepaspectratio]{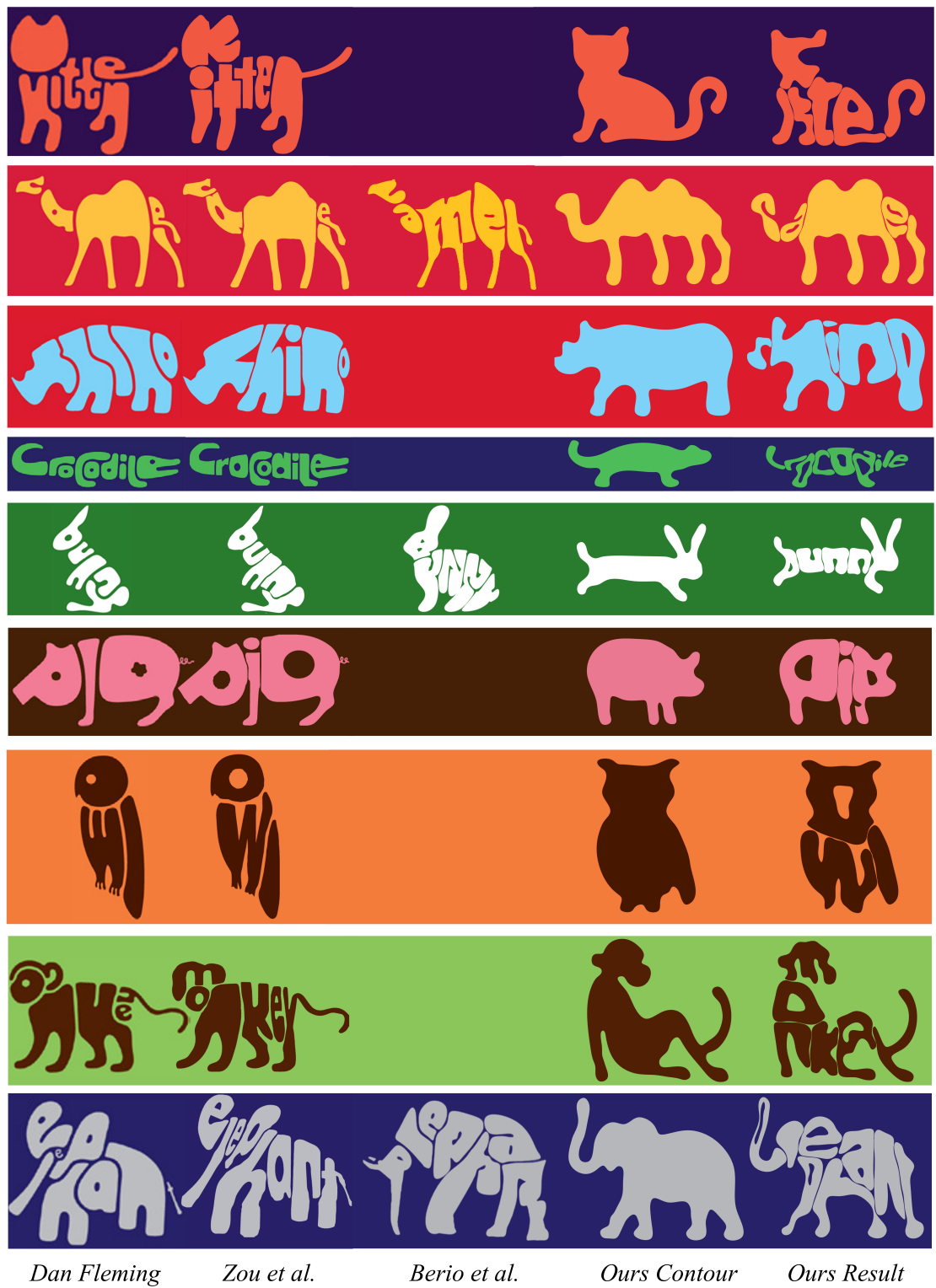}
  \caption{Comparison with methods that generate calligrams from a given image or contour. The human-designed examples by Dan Fleming and the automatic results of Zou et al. are reproduced from \emph{Legible Compact Calligrams}~\cite{zou2016legible}, and the B-spline-based results of Berio et al. are from \emph{Neural Image Abstraction Using Long Smoothing B-Splines}~\cite{berio2025bsplines}. The rightmost examples are generated by our method from text alone. Blank entries indicate cases that were not reported in the original source.}
  \label{fig:comparisons}
\end{figure*}

\section{Comparison and Results}

\subsection{Comparison with Prior Methods}

We compare Telligram with contour-conditioned calligram methods, including the method of Zou et al.~\cite{zou2016legible} and the B-spline-based abstraction method of Berio et al.~\cite{berio2025bsplines}. These methods require a given image or contour to define the target outer shape, whereas our method starts from text alone. As shown in Fig.~\ref{fig:comparisons}, this difference lets our method form the semantic outer structure and letter placement jointly, so the resulting branches are more naturally supported by glyph mass instead of being stretched only to satisfy a prescribed outline. We do not report direct quantitative comparison with these two methods because neither paper provides official code, so in practice we can only perform qualitative comparison using the images reported in their papers.

\begin{figure}[t]
  \centering
  \includegraphics[width=0.6\columnwidth]{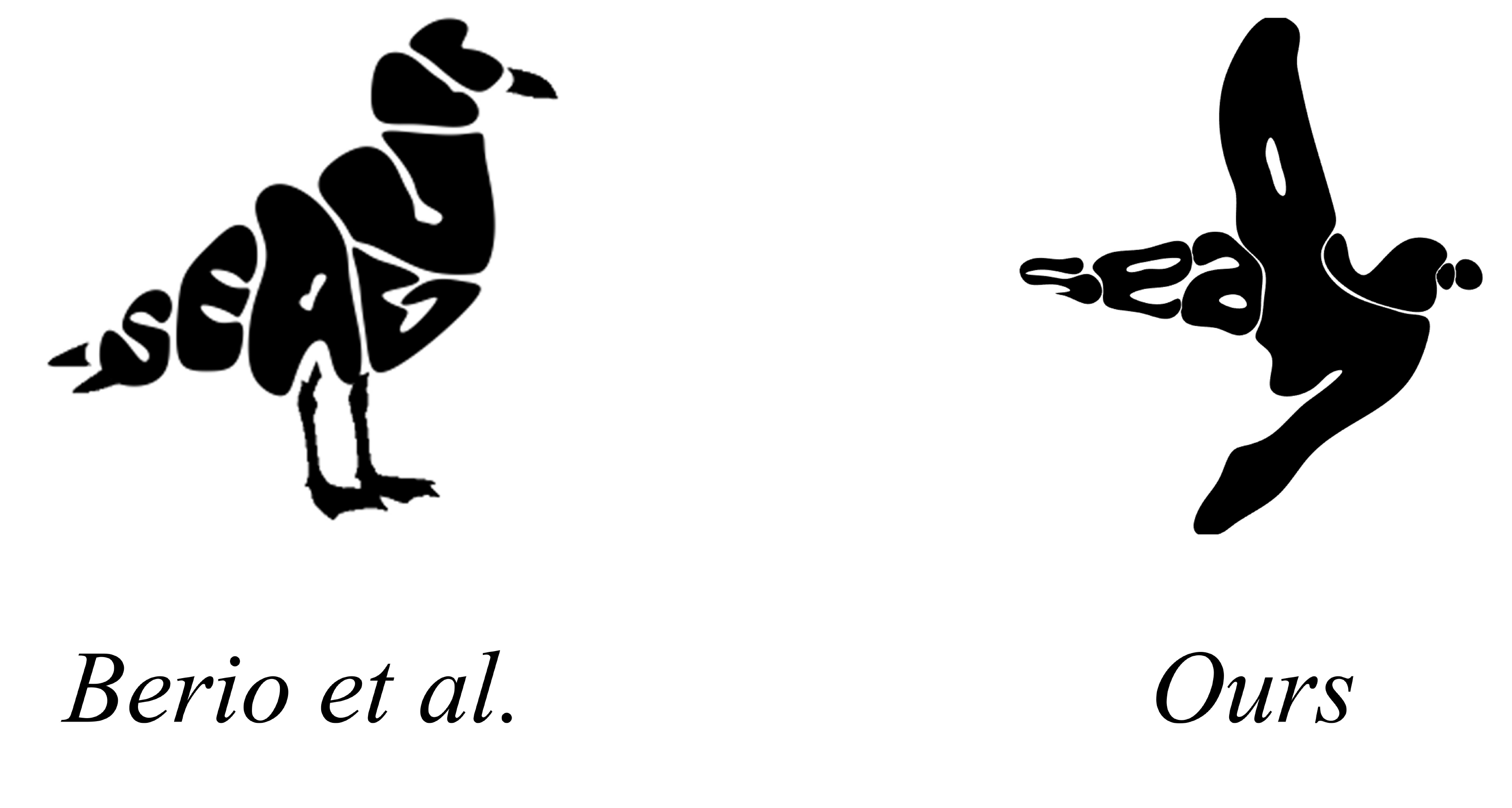}
  \caption{Comparison between direct silhouette prompting and skeleton-guided semantic prior formation. The left examples are generated by Berio et al.~\cite{berio2025bsplines} from prompts such as ``silhouette of a seagull'', and the right examples are produced by our method under the same semantic target.}
  \label{fig:comparisons2}
\end{figure}

\begin{figure}[t]
  \centering
  \includegraphics[width=\columnwidth]{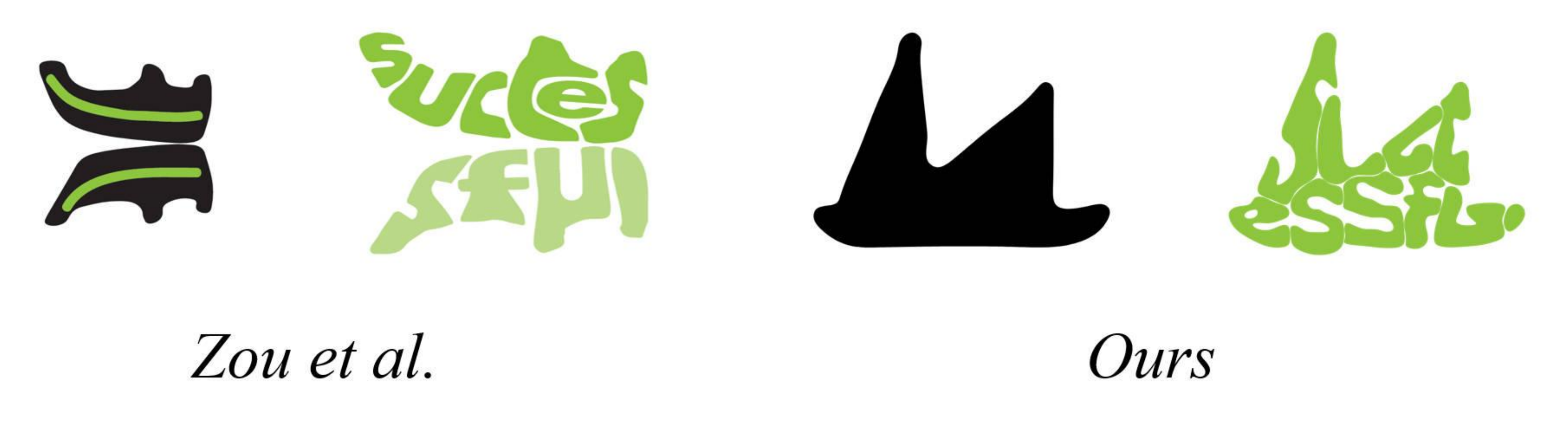}
  \caption{Comparison under matched semantic information, namely the concept ``boat'', and the same word ``successful''. The left examples are from Zou et al.~\cite{zou2016legible}, and the right examples are produced by our method. For the Zou et al. example, the leftmost black silhouette is the contour image shown together with their result in the original paper.}
  \label{fig:comparisons4}
\end{figure}

A second comparison shows the role of skeleton-guided semantic prior formation. Berio et al.~\cite{berio2025bsplines} directly prompt the model with descriptions such as ``silhouette of a seagull''. This can produce a semantically recognizable outer contour, but it may also create thin branches that are hard to fill with readable glyphs. As shown in Fig.~\ref{fig:comparisons2}, our method uses letter skeletons as a structural prior when forming the semantic occupancy prior. This makes it easier to preserve semantic contour quality while reducing branches that are too thin to be occupied by letters in a reasonable way.

A third comparison uses matched semantic information, namely the concept ``boat'', together with the same word ``successful''. As shown in Fig.~\ref{fig:comparisons4}, Zou et al.~\cite{zou2016legible} require explicit guiding lines to arrange multiple text lines, whereas our method starts from a multi-line text layout and forms the semantic figure through the same semantic occupancy optimization.

\subsection{Additional Qualitative Results}

We further show qualitative results on different prompts and word structures. These examples illustrate the visual diversity of the generated calligrams and show that the method can produce recognizable figurative layouts without an input contour image.

\begin{figure}[t]
  \centering
  \includegraphics[width=\columnwidth]{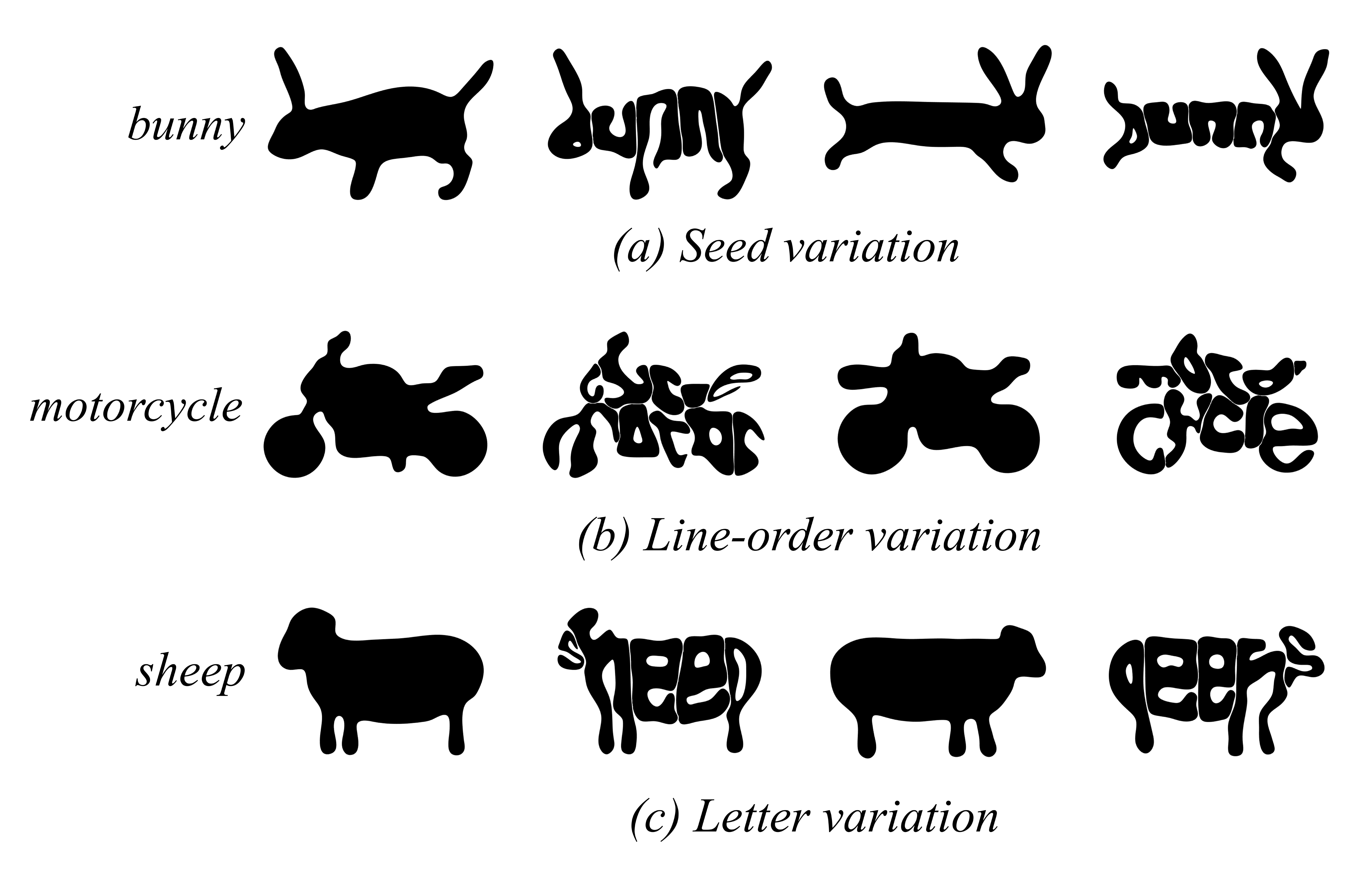}
  \caption{Examples of output diversity. (a) shows results generated from the same prompt, word, and layout under different random seeds. (b) shows line-order variation, namely different line arrangements of the same word such as exchanging the upper and lower line while preserving the same set of letters. (c) shows results of the same letters arranged in different orders under the same prompt.}
  \label{fig:comparisons3}
\end{figure}

\begin{figure*}[t]
  \centering
  \includegraphics[width=\textwidth]{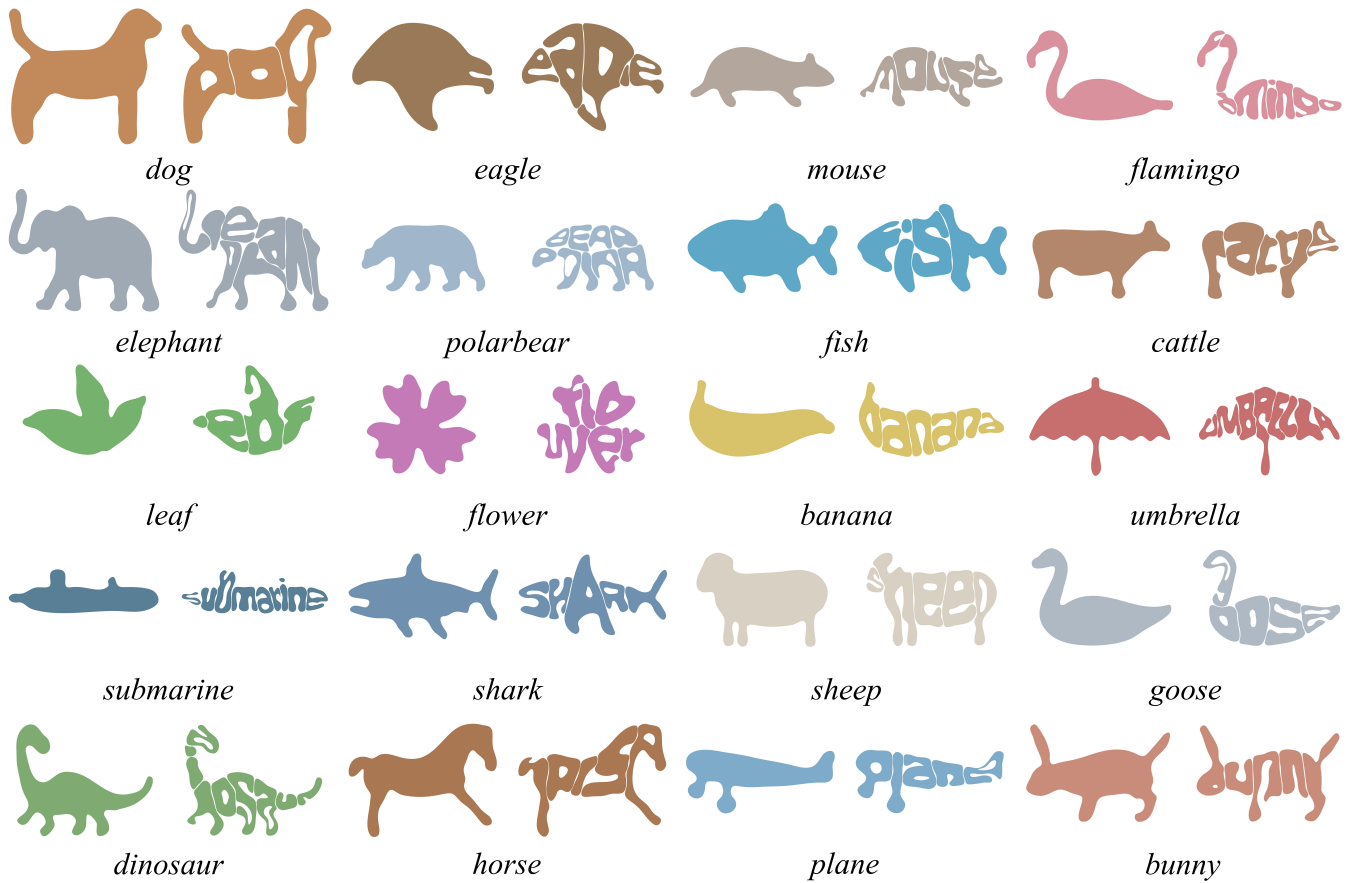}
  \caption{Additional qualitative results of Telligram under diverse prompts. Across different semantic targets, the method produces creative and varied calligram layouts while maintaining recognizable figurative structure.}
  \label{fig:all_results}
\end{figure*}

Figure~\ref{fig:comparisons3} shows controlled variations under different random seeds, different line orders, and different words under the same prompt. Here, line-order variation refers to rearranging the reading lines of a multi-line word while keeping the same letter set, for example, by changing the order of the top and bottom lines. Beyond these controlled settings, Fig.~\ref{fig:all_results} demonstrates broader prompt coverage. Across a wide range of semantic prompts, Telligram produces varied calligram layouts without collapsing to a single repeated template.

\section{Conclusion}

This paper presented Telligram, a training-free two-stage framework for text-only calligram generation. The method first builds a semantic occupancy prior under diffusion guidance and then converts this prior into readable final glyphs through lightweight geometric processing. The experiments showed that hierarchical gradient projection provides better semantic alignment than direct joint update or smoothed joint update, while avoiding the heavy tuning burden of manually staged schedules.

The current framework also has clear limitations. Because the representation is based on silhouettes and skeletons with finite thickness, it can lose fine-grained appearance cues and has limited ability to restore letters once severe deformation has already occurred. The current system also uses DeepFloyd guidance together with a low semantic resolution of $96\times 96$, so different prompts may sometimes produce similar global shapes.

\begin{figure}[t]
  \centering
  \includegraphics[width=\columnwidth]{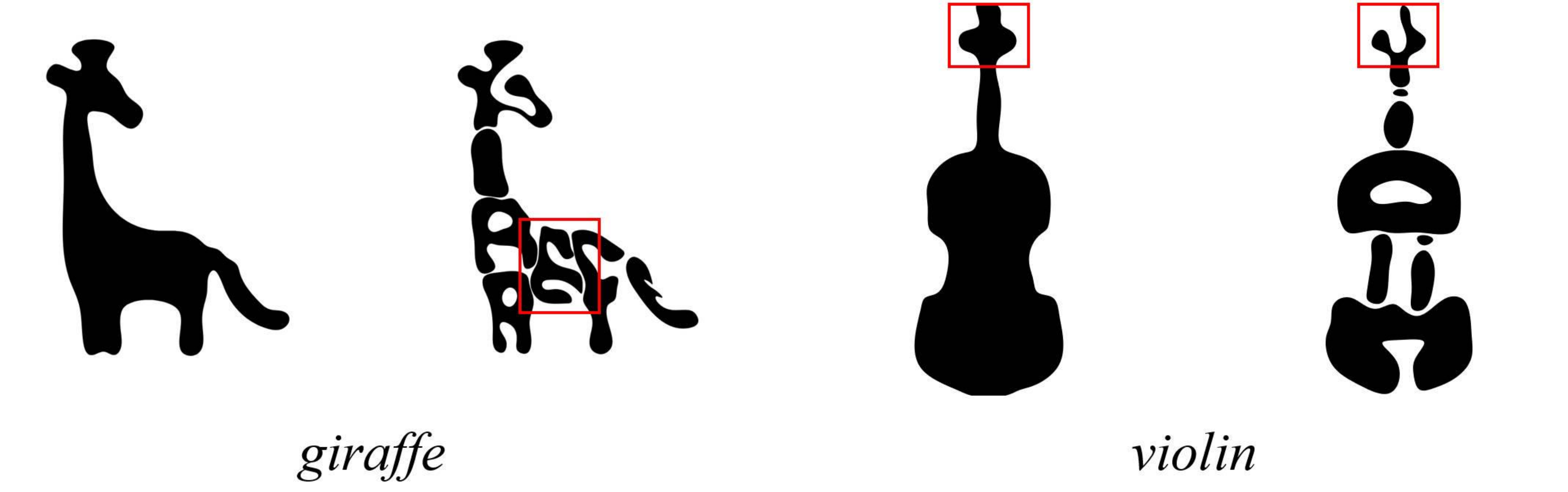}
  \caption{Representative failure cases. Left: the giraffe silhouette remains semantically related but may still be recognized as a nearby category such as \emph{camel}, and severe intermediate deformation can weaken the intended letter identity, as highlighted in the red box. Right: even when the coarse letter shape is restored, branch-like support may still persist in the final glyph, as highlighted in the \emph{v} of the violin example.}
  \label{fig:failure_case}
\end{figure}

Fig.~\ref{fig:failure_case} illustrates three representative failure modes in the current training-free setting. First, silhouette-like outputs may remain semantically related while still being ambiguous under category recognition; for example, the giraffe silhouette may be recognized as \emph{camel} in CLIP zero-shot top-1 prediction. On a balanced test set of 36 target categories with 15 random seeds per category (540 generated samples in total), we evaluate semantic recognition using CLIP zero-shot classification over a combined label space consisting of ImageNet-1k labels and our target labels. Under a top-5 check, this gives a semantic recognition rate of 72.03\%. Here, a top-5 semantic match means that the predictions contain either the target category itself or a more specific label with the same target noun, such as \emph{bear} matched by \emph{grey bear}; visually similar but semantically different categories are not counted as matches. This number is used only as a coarse category-level indicator for silhouette-like outputs. Second, hierarchical gradient projection reduces instability caused by entangled coarse and local deformation during Stage 1, but it does not fully prevent failures driven mainly by local identity-damaging skeleton updates. If a letter identity is already heavily weakened before entering Stage 2, the later convex-hull-based restoration may still be insufficient to recover the intended glyph structure, as illustrated by the highlighted region in the giraffe example. Third, even when Stage 2 restores the coarse letter shape, branch-like support may still persist in the final glyph, as in the highlighted \emph{v} of the violin example.

These limitations are tied in part to our decision to keep the pipeline training-free. A promising future direction is therefore to relax this constraint and learn a task-specific readability guider or letter-identity discriminator that provides stable identity-preserving gradients already in Stage 1. Such a model could reduce severe intermediate deformation before Stage 2, alleviate later restoration failures, and support a more tightly coupled formulation of semantic shape formation and readability preservation. Another possible extension is to enrich the current geometric hierarchy with additional deformation variables for finer boundary refinement once stable identity-preserving guidance becomes available.

\section*{Acknowledgments}
We thank the anonymous reviewers for their constructive feedback. This work was supported by the National Natural Science Foundation of China (62472287) and the Natural Science Foundation of Shenzhen City (JCYJ20250604181519025).

\bibliographystyle{eg-alpha-doi}
\bibliography{sections/references}

@String{tog = "ACM TOG"}

@inproceedings{xu2007calligraphic,
  author    = {Jie Xu and Craig S. Kaplan},
  title     = {Calligraphic Packing},
  booktitle = {Graphics Interface},
  year      = {2007},
  pages     = {43--50},
  publisher = {Canadian Human-Computer Communications Society},
}

@article{zou2016legible,
  author     = {Changqing Zou and Junjie Cao and Warunika Ranaweera and Ibraheem Alhashim and Ping Tan and Alla Sheffer and Hao Zhang},
  title      = {Legible Compact Calligrams},
  journal    = {ACM TOG},
  year       = {2016},
  volume     = {35},
  number     = {4},
  articleno  = {122},
}

@inproceedings{shao2026score,
  author    = {Zefan Shao and Jin Zhou and Hongliang Yang and Pengfei Xu},
  title     = {SCORE: Semantic Collage by Optimizing Rendered Elements},
  booktitle = {AAAI},
  year      = {2026},
  pages     = {2038--2046},
}

@article{zhang2017ornamental,
  author    = {Junsong Zhang and Yu Wang and Weiyi Xiao and Zhenshan Luo},
  title     = {Synthesizing Ornamental Typefaces},
  journal   = {CGF},
  year      = {2017},
  volume    = {36},
  number    = {1},
  pages     = {64--75},
}

@article{zhang2022wordpaintings,
  author     = {Junsong Zhang and Zuyi Yang and Linchengyu Jin and Zhitang Lu and Jinhui Yu},
  title      = {Creating Word Paintings Jointly Considering Semantics, Attention, and Aesthetics},
  journal    = {ACM TAP},
  year       = {2022},
  volume     = {19},
  number     = {3},
  articleno  = {13},
  pages      = {1--21},
}

@article{iluz2023wordasimage,
  author    = {Shir Iluz and Yael Vinker and Amir Hertz and Daniel Berio and Daniel Cohen{-}Or and Ariel Shamir},
  title     = {Word-As-Image for Semantic Typography},
  journal   = {ACM TOG},
  year      = {2023},
  volume    = {42},
  number    = {4},
  pages     = {1--11},
}

@inproceedings{tanveer2023dsfusion,
  author       = {Maham Tanveer and Yizhi Wang and Ali Mahdavi{-}Amiri and Hao Zhang},
  title        = {DS-Fusion: Artistic Typography via Discriminated and Stylized Diffusion},
  booktitle    = {ICCV},
  year         = {2023},
  pages        = {374--384},
}

@inproceedings{wang2023anythingtoglyph,
  author    = {Changshuo Wang and Lei Wu and Xiaole Liu and Xiang Li and Lei Meng and Xiangxu Meng},
  title     = {Anything to Glyph: Artistic Font Synthesis via Text-to-Image Diffusion Model},
  booktitle = {SIGGRAPH Asia},
  year      = {2023},
  pages     = {1--11},
  publisher = {ACM},
}

@inproceedings{hussein2024khattat,
  author       = {Ahmed Hussein and Alaa Elsetohy and Sama Hadhoud and Tameem Bakr and Yasser Rohaim and Badr AlKhamissi},
  title        = {Khattat: Enhancing Readability and Concept Representation of Semantic Typography},
  booktitle    = {ECCV Workshops},
  year         = {2025},
  pages        = {278--295},
  publisher    = {Springer},
}

@article{tatsukawa2024fontclip,
  author    = {Yuki Tatsukawa and I{-}Chao Shen and Anran Qi and Yuki Koyama and Takeo Igarashi and Ariel Shamir},
  title     = {FontCLIP: A Semantic Typography Visual-Language Model for Multilingual Font Applications},
  journal   = {CGF},
  year      = {2024},
  volume    = {43},
  number    = {2},
  pages     = {e15043},
}

@article{wang2021deepvecfont,
  author    = {Yizhi Wang and Zhouhui Lian},
  title     = {DeepVecFont: Synthesizing High-Quality Vector Fonts via Dual-Modality Learning},
  journal   = {ACM TOG},
  year      = {2021},
  volume    = {40},
  number    = {6},
}

@inproceedings{xia2023vecfontsdf,
  author    = {Zeqing Xia and Bojun Xiong and Zhouhui Lian},
  title     = {VecFontSDF: Learning To Reconstruct and Synthesize High-Quality Vector Fonts via Signed Distance Functions},
  booktitle = {CVPR},
  year      = {2023},
  pages     = {1848--1857},
}

@inproceedings{thamizharasan2024vecfusion,
  author    = {Vikas Thamizharasan and Difan Liu and Shantanu Agarwal and Matthew Fisher and Micha{"e}l Gharbi and Oliver Wang and Alec Jacobson and Evangelos Kalogerakis},
  title     = {VecFusion: Vector Font Generation with Diffusion},
  booktitle = {CVPR},
  year      = {2024},
  pages     = {7943--7952},
}

@inproceedings{wang2022textlogo,
  author    = {Yizhi Wang and Guo Pu and Wenhan Luo and Yexin Wang and Pengfei Xiong and Hongwen Kang and Zhouhui Lian},
  title     = {Aesthetic Text Logo Synthesis via Content-Aware Layout Inferring},
  booktitle = {CVPR},
  year      = {2022},
  pages     = {2436--2445},
}

@article{berio2025bsplines,
  author    = {Daniel Berio and Michael Stroh and Sylvain Calinon and Frederic Fol Leymarie and Oliver Deussen and Ariel Shamir},
  title     = {Neural Image Abstraction Using Long Smoothing B-Splines},
  journal   = {ACM TOG},
  year      = {2025},
  volume    = {44},
  number    = {6},
  pages     = {1--11},
}

@inproceedings{cui2010contextwordcloud,
  author    = {Weiwei Cui and Yingcai Wu and Shixia Liu and Furu Wei and Michelle X. Zhou and Huamin Qu},
  title     = {Context Preserving Dynamic Word Cloud Visualization},
  booktitle = {PacificVis},
  year      = {2010},
  pages     = {121--128},
}

@article{koh2010maniwordle,
  author    = {Kyle Koh and Bongshin Lee and Bohyoung Kim and Jinwook Seo},
  title     = {ManiWordle: Providing Flexible Control over Wordle},
  journal   = {IEEE TVCG},
  year      = {2010},
  volume    = {16},
  number    = {6},
  pages     = {1190--1197},
}

@article{wang2020shapewordle,
  author    = {Yunhai Wang and Xiaowei Chu and Kaiyi Zhang and Chen Bao and Xiaotong Li and Jian Zhang and Chi{-}Wing Fu and Christophe Hurter and Bongshin Lee and Oliver Deussen},
  title     = {ShapeWordle: Tailoring Wordles using Shape-aware Archimedean Spirals},
  journal   = {IEEE TVCG},
  year      = {2020},
  volume    = {26},
  number    = {1},
  pages     = {991--1000},
}

@inproceedings{waldner2013facetclouds,
  author    = {Manuela Waldner and Johann Schrammel and Michael Klein and Katrin Kristjansdottir and Dominik Unger and Manfred Tscheligi},
  title     = {FacetClouds: Exploring Tag Clouds for Multi-Dimensional Data},
  booktitle = {Graphics Interface},
  year      = {2013},
  pages     = {17--24}
}

@inproceedings{kaser2007tagcloud,
  author       = {Owen Kaser and Daniel Lemire},
  title        = {Tag-Cloud Drawing: Algorithms for Cloud Visualization},
  booktitle    = {Tagging and Metadata for Social Information Organization (WWW 2007 Workshop)},
  year         = {2007},
}

@inproceedings{poole2022dreamfusion,
  author    = {Ben Poole and Ajay Jain and Jonathan T. Barron and Ben Mildenhall},
  title     = {DreamFusion: Text-to-3D using 2D Diffusion},
  booktitle = {ICLR},
  year      = {2023},
}

@inproceedings{wang2023prolificdreamer,
  author    = {Zhengyi Wang and Cheng Lu and Yikai Wang and Fan Bao and Chongxuan Li and Hang Su and Jun Zhu},
  title     = {ProlificDreamer: High-Fidelity and Diverse Text-to-3D Generation with Variational Score Distillation},
  booktitle = {NeurIPS},
  year      = {2023},
}

@article{maharik2011digitalmicrography,
  author    = {Ron Maharik and Mikhail Bessmeltsev and Alla Sheffer and Ariel Shamir and Nathan Carr},
  title     = {Digital Micrography},
  journal   = {ACM TOG},
  year      = {2011},
  volume    = {30},
  number    = {4},
  articleno = {100},
  pages     = {100:1--100:12},
}

@misc{wikipedia2014calligram,
  author       = {{Wikipedia contributors}},
  title        = {Calligram --- Wikipedia{,} The Free Encyclopedia},
  year         = {2026},
  url          = {https://en.wikipedia.org/wiki/Calligram},
  note         = {Accessed: 2026-05-28}
}

@inproceedings{radford2021clip,
  author    = {Alec Radford and Jong Wook Kim and Chris Hallacy and Aditya Ramesh and Gabriel Goh and Sandhini Agarwal and Girish Sastry and Amanda Askell and Pamela Mishkin and Jack Clark and Gretchen Krueger and Ilya Sutskever},
  title     = {Learning Transferable Visual Models From Natural Language Supervision},
  booktitle = {ICML},
  series    = {Proceedings of Machine Learning Research},
  volume    = {139},
  year      = {2021},
  pages     = {8748--8763},
}

@misc{deepfloyd2023if,
  author       = {{DeepFloyd} and {Stability AI}},
  title        = {IF-I-M-v1.0 Model Card},
  year         = {2023},
  howpublished = {Hugging Face model card for DeepFloyd/IF-I-M-v1.0},
  url          = {https://huggingface.co/DeepFloyd/IF-I-M-v1.0},
  note         = {Stage-I pixel diffusion model with frozen T5 text encoder}
}

@inproceedings{li2021trocr,
  author       = {Minghao Li and Tengchao Lv and Jingye Chen and Lei Cui and Yijuan Lu and Dinei Florencio and Cha Zhang and Zhoujun Li and Furu Wei},
  title        = {TrOCR: Transformer-based Optical Character Recognition with Pre-trained Models},
  booktitle    = {AAAI},
  year         = {2023},
  pages        = {13094--13102},
}

@article{shi2015crnn,
  author       = {Baoguang Shi and Xiang Bai and Cong Yao},
  title        = {An End-to-End Trainable Neural Network for Image-Based Sequence Recognition and Its Application to Scene Text Recognition},
  journal      = {IEEE TPAMI},
  year         = {2017},
  volume       = {39},
  number       = {11},
  pages        = {2298--2304},
}

@inproceedings{bautista2022parseq,
  author       = {Darwin Bautista and Rowel Atienza},
  title        = {Scene Text Recognition with Permuted Autoregressive Sequence Models},
  booktitle    = {ECCV},
  year         = {2022},
  pages        = {178--196},
  publisher    = {Springer},
}

@misc{openai2026gptimage2,
  author       = {{OpenAI}},
  title        = {Introducing ChatGPT Images 2.0},
  year         = {2026},
  url          = {https://openai.com/index/introducing-chatgpt-images-2-0/},
  note         = {Accessed: 2026-06-09}
}

\end{document}